\documentclass[twocolumn]{cinc}
\usepackage{graphicx}
\usepackage{amsmath}
\begin{document}
\bibliographystyle{cinc}

\title{Spectral Distribution Complexity of the Surface Fibrillatory Waves Predicts Post-Catheter Ablation Relapse in Persistent Atrial Fibrillation}


\author {Pilar Escribano$^{1}$, Juan Ródenas$^{1}$, Manuel García$^{1}$, Miguel A Arias$^{2}$, José J Rieta$^{3}$, Raúl Alcaraz$^{1}$ \\
\ \\ 
$^1$ Research Group in Electronic, Biomedical and Telecomm. Eng., Univ. of Castilla-La Mancha, Spain\\
$^2$ Cardiac Arrhythmia Department, Hospital Virgen de la Salud, Toledo, Spain\\
$^3$ BioMIT.org, Electronic Engineering Department, Universistat Politecnica de Valencia, Spain}

\maketitle

%
%
%
%
\begin{abstract}
	As for most of cardiac arrhythmias, atrial fibrillation (AF) is primarily treated by catheter ablation (CA). However, the mid-term recurrence rate of this procedure in persistent AF patients is still limited and the preoperative prediction of its outcome is clinically interesting to select candidates who could benefit the most from the intervention. This context encouraged the study of $C_0$ complexity as a novel predictor, because it estimates organization of the power spectral distribution (PSD) of the fibrillatory waves ($f$-waves). For that purpose, the PSD was divided into two divergent components using a threshold, $\theta$, which was considered by multiplying the mean value of the PSD by a factor, $\alpha$, ranging between 1.5 and 2.5. On a database of 74 patients, the values of $C_0$ complexity computed for all $\alpha$ factors reported statistically significant differences between the patients who maintained sinus rhythm and those who relapsed to AF after a follow-up of 9 months. They also showed higher values of sensitivity (Se), specificity (Sp), and accuracy (Acc) than the well known predictors of the dominant frequency (DF) and $f$-wave amplitude. Moreover, the combination of the DF and the $C_0$ complexity computed with $\alpha =2$, via a decision tree, improved classification until values of Se, Sp and Acc of 75.33, 77.33 and 76.58\%, respectively. These results manifests the relevance of the $f$-wave PSD distribution to anticipate CA outcome in persistent AF patients.
\end{abstract}
\section{Introduction}
	Atrial fibrillation (AF) is a type of supraventricular tachyarrhytmia that has become one of the most important public health issues~\cite{Zoni2014}, reaching a prevalence of between 2 and 4\% in adults, which is expected to continue increasing~\cite{Hindricks:2021aa}. It is associated with symptoms that severely affect the quality of life of patients, such as palpitations attacks and shortness of breath, among others, and is the most common risk factor of ischemic stroke~\cite{Warmus2020}. AF leads to an atrial substrate remodelling process that favors the perpetuation of the arrhythmia~\cite{Schotten:2016aa}, so that it should be diagnosed and treated as soon as possible~\cite{Nattel:2014aa}.
	
	Catheter ablation (CA) is a well-established treatment for AF patients that is very suitable to end the arrhythmia related-symptoms, as it aims at restoring sinus rhythm (SR)~\cite{Hindricks:2021aa}. However, persistent AF patients, who are those suffering from arrhythmic episodes that do not self-terminate within a week and require external intervention for their termination, present a high mid-term recurrence rate (about 35\% within the first year), despite applying additional ablation lines or repeated procedures~\cite{Schmidt2020}. This situation has generated clinical interest in preoperative prediction of CA result to select those candidates really suitable for this treatment and, therefore, avoid unnecessary risk, reduce hospitalization rates and repeated procedures, and minimize the health care system cost and burden associated with AF treatment~\cite{Walsh2018}.
	
	So far, clinical predictors addressed by previous works have only provided limited and controversy results~\cite{Dretzke:2020aa}. Alternatively, some electrocardiographic (ECG)-based markers of the fibrillatory waves ($f$-waves), such as the dominant frequency (DF)~\cite{Alcaraz2016} and $f$-wave amplitude (FWA)~\cite{Nault:2009aa}, have also reported a promising ability to anticipate CA result. However, they are still insufficient for clinical application and, hence, the aim of this work is to explore a novel $f$-wave frequency domain metric to improve preoperative prediction of CA outcome.

\section{Materials and methods}
\subsection{Study population}
	In the study, 74 persistent AF patients (21 women and 53 men), with mean age of 65 years (between 39 and 82 years) were enrolled. All were targeted under standard clinical indications to their first radio-frequency CA intervention at the University Hospital of Toledo, Spain.
	
\subsection{CA protocol and further follow-up}
	CA avoids open heart surgery by inserting catheters through the femoral venous access, while the patient is sedated and is under anticoagulant drugs. It is mainly based on pulmonary veins isolation (PVI) technique that aims to end the abnormal electrical activity triggered in that region, which is accessed by transseptal puncture. This was carried out by delivering point-by-point ablative lesions with a radio-frequency catheter to create a contiguous antral circumferential line around each pulmonary vein (PV), whose location was determined by a mapping catheter \cite{Morin2016}. Finally, the end of the procedure was achieved when all PVs were successfully isolated or, if needed,  after restoring SR by electrical cardioversion. All interventions were successfully completed and patients did not suffer from any complication. After a follow-up period of nine months, during which patients received anticoagulants and antiarrhythmic drugs at clinical judgment, 19 patients relapsed to AF and the remaining 55 maintained SR.
	
\subsection{Signal acquisition and preprocessing}
	The heart rhythm of the patients was preoperatively recorded through a standard 12-lead ECG signal with a duration between 6 seconds and 5 minutes. These signals, together with the clinical information after the follow-up period, constituted the database of the study. Signal acquisition was performed with a 977~Hz sampling rate and 16 bits resolution. Afterwards, lead V1 was selected for the extraction of the $f$-waves, because it usually presents the greatest amplitude with respect to the ventricular activity~\cite{Petr_nas_2018}. Previously, this lead was preprocessed by common filtering and wavelet-based denoising to attenuate baseline wandering, powerline interference, and high frequency noise components~\cite{Leif2005,Garcia:2018aa}. 
	The $f$-waves were finally extracted using an algorithm based on an adaptive singular value cancellation of QRST complexes~\cite{Alcaraz:2008aa}.
	  
\subsection{Characterization of the $f$-waves}
Given the disparity in the ECG length acquired from the patients, the extracted $f$-wave signal from each lead V1 was chopped into non-overlapped 6 second-length intervals. The maximum number of segments per patient was limited to 5 and the ECG-derived parameters from each one were averaged for unbiased subject-based analysis and classification. Regarding the characterization of each 6 second-length interval, the DF and the normalized FWA (nFWA) were evaluated as reference parameters, since they were broadly analysed in previous works. The frequency metric was computed from the power spectral density (PSD) of each excerpt, estimated through a Welch Periodogram with a 4,000 point-length Hamming window, 3,000 points of overlapping between adjacent windowed sections, and a spectral resolution of 0.1~Hz. The DF was considered as the highest PSD peak within the 3--25 Hz range~\cite{Alcaraz2016}. On the other hand, the FWA was estimated as the root mean square value of the $f$-waves in the time domain~\cite{Alcaraz:2011aa} and the nFWA as a percentage of this FWA regarding the median amplitude of the R-peaks to minimize the influence of spurious effects that could alter ECG amplitude, such as the skin conductivity.
	
Additionally, a novel parameter, i.e. the $C_0$ complexity, was obtained as a non-linear measure of the spectral organization of the $f$-waves~\cite{Yueli2008}. It divides the PSD of the signal into two divergent components, a regular and an irregular part. Its computation was carried out after normalizing the PSD over the total power of the frequency band between 3 and 25 Hz to obtain a unit area probability function, i.e., 
\begin{equation}
	PSD_n\left(f\right) = \frac{PSD\left(f\right)}{\sum_{f_l}^{f_u}PSD\left(f\right)}.
\end{equation}
	
\noindent Different threshold values, $\theta$, to separate both parts of the PSD were then evaluated by multiplying its mean value to a factor, $\alpha$, ranging between 1.5 to 2.5. In this way, the irregular part was obtained as the PSD below the computed threshold, being the upper region the regular part. Precisely, the values higher than $\theta$ were rejected by substituting them with zero to obtain the irregular part, i.e, 
\begin{equation}
	PSD(f)^{\prime} = \left\{ \begin{matrix}
		PSD(f), \ \ \ if \ PSD(f) \leq \theta, \\\
	0,\ \ \ \ \ \ \ \ \ \ \ \ \ \ \ if \ PSD(f) > \theta,
	\end{matrix}\right.
	\end{equation}

	Finally, the $C_0$ complexity was computed as the ratio of the irregular part to the total PSD distribution, i.e., 
		\begin{equation}
		C_0 = \frac{\sum_{f_l}^{f_u}PSD_n^\prime\left(f\right)}{\sum_{f_l}^{f_u}PSD_n\left(f\right)},
		\end{equation}
	where $f_l$ and $f_u$ were the lower and upper frequency band limits, 3 and 25 Hz, respectively. This index then results in a real number between 0 and 1, so that the greater the predomination of the irregular part of the signal, the higher the value of $C_0$ ~\cite{Yueli2008}.  
		
\subsection{Performance assessment}
The analyzed metrics were expressed in terms of median and interquartile range (IQR) for patients who maintained SR and relapsed to AF. Indeed, Lilliefors and Levene's tests provided that most of data distributions were non-normal but homocedastic. A non-parametric Mann Whitney U test was also used to evaluate statistical differences between both groups of patients.
\par
A hold-out classification approach with 2/3 of the database for training and 1/3 for validation was repeated 100 times to estimate the prediction performance of each single index. In the training phase, a receiver operating characteristic (ROC) curve was employed to estimate optimal threshold to discern between the two groups of patients. This tool plots the fraction of true positives out of positives (sensitivity) against the fraction of false positives out of negatives (1$-$specificity) at various threshold settings, considering sensitivity (Se) and specificity (Sp) as the rate of patients relapsing to AF and maintaining SR correctly classified, respectively. The optimal threshold was selected as that providing the best balance between Se and Sp, although this may not reach the highest accuracy (Acc), which was the percentage of patients correctly classified. Finally, the area under the ROC curve (AUC) was also computed as an aggregate measure of performance of a variable across all possible classification thresholds. 
\par
To explore complementary information among single features and improve prediction of CA outcome, a multivariate analysis based on a decision tree model of 5 splits was also conducted. A forward sequential feature selection algorithm minimizing the classification error addressed the DF and $C_0$ complexity with  $\alpha$ = 2 as the best metrics to built a prediction model, whose classification performance was assessed by 100 cycles as single indices.

\section{Results}
Table 1 shows the values obtained by the single indices for the two groups of patients. As can be seen, all reported statistically significant differences, obtaining $p$-values lower than 0.05, except for the case of nFWA. Moreover, whereas the DF and nFWA followed the same trend of previous works with higher and lower values for those patients who relapsed to AF than for those who maintained SR, respectively, $C_0$ complexity provided lower values for the first group regardless of the threshold $\theta$. 
		\begin{table}[!b]
		\caption{Values and statistical differentiation of the analyzed metrics for the two groups of patients.}
		\vspace{2 mm}
		\centering
		\begin{tabular}{@{\extracolsep{-6pt}}lccc}
			\hline\hline
			& \multicolumn{2}{c}{Group of patients who} &\\
			\cline{2-3}
			Index 						& maintained SR		& relapsed to AF 		& $p$-value \\
			\hline 
			DF (Hz)						&\( 5.34\ (1.44) \)		&\( 5.68\ (1.47) \)		&\( 0.036 \)\\
			nFWA (\%)					&\( 6.41\ (7.21) \)		&\( 5.58\ (4.10) \)		&\( 0.316 \)\\
			$C_0$ [$\alpha$=$1.5]$(\%)		&\( 26.41\ (7.24) \)	&\( 23.25\ (3.20) \)	&\( 0.004 \)\\
			$C_0$ [$\alpha$=$1.75]$(\%)	&\( 29.15\ (6.98) \)	&\( 25.44 (2.97) \)		&\( 0.002 \)\\
			$C_0$ [$\alpha$=$2]$(\%)		&\( 31.79\ (8.49) \)	&\( 26.72\ (3.13) \)	&\( <0.001 \)\\
			$C_0$ [$\alpha$=$2.25]$(\%)	&\( 33.70\ (9.33) \)	&\( 28.60\ (4.26) \)	&\( 0.003 \)\\
			$C_0$ [$\alpha$=$2.5]$(\%) 	&\( 35.94\ (10.20) \)	&\( 31.25\ (3.34) \)	&\( 0.004 \)\\
			\hline\hline    
		\end{tabular}
		\label{tab:table1}
	\end{table}	
	\begin{table}[!b]
		\caption{Classification results obtained by the metrics computed from the $f$-waves.}
		\centering
		\vspace{2 mm}
		\begin{tabular}{lcccc}
			\hline\hline
			Index 					& Se (\%) 	& Sp (\%)	& Acc (\%)	& AUC (\%) \\
			\hline
			DF  					&\(63.16\)	&\(61.82\)	&\(62.16\)	&\(66.22\)\\
			nFWA 					&\(52.63\)	&\(54.55\)	&\(54.05\)	&\(57.80\)\\
			$C_0$[$\alpha$=$1.5]$  	&\(68.42\)	&\(69.09\)	&\(68.92\)	&\(72.44\)\\
			$C_0$[$\alpha$=$1.75]$		&\(73.68\)	&\(72.73\)	&\(72.97\)	&\(74.45\)\\
			$C_0$[$\alpha$=$2]$  		&\(68.42\)	&\(70.91\)	&\(70.27\)	&\(75.50\)\\
			$C_0$[$\alpha$=$2.25]$ 	&\(68.42\)	&\(70.91\)	&\(70.27\)	&\(72.73\)\\
			$C_0$[$\alpha$=$2.5]$ 		&\(68.42\)	&\(69.09\)	&\(68.92\)	&\(72.15\)\\
			\hline\hline    
		\end{tabular}
		\label{tab:table2}
	\end{table}

On the other hand, classification performance of each individual parameter is presented by Table 2. The highest discriminant ability was achieved by the $C_0$ complexity computed with $\alpha$ = 1.75, although the greatest AUC was obtained when $\alpha$ = 2. Compared to these indices, the combination of the DF and $C_0$ complexity computed with $\alpha$ = 2, via a classification tree, reached improvements of up to 23\%, since values of Se, Sp, Acc and AUC of 75.33, 77.33, 76.58, and 81.00\% were obtained, respectively.

\section{Discussion and conclusions} 
The high mid-term recurrence rate of persistent AF patients undergoing CA places a high burden in health care systems, thus increasing clinical interest in preoperative prediction of the procedure outcome, since it remains the first-line treatment for AF~\cite{Zoni2014}. Knowing success probability before the intervention would allow optimal selection of patients, avoiding unnecessary risks and minimizing the health care costs in AF treatment, among other benefits \cite{Walsh2018}. For that prediction, $C_0$ complexity of the $f$-waves has been introduced in the present work. Regardless of the factor $\alpha$, this index has exhibited better classification results than the common predictors of DF and nFWA. Moreover, the combination of this index, computed for $\alpha$ = 2, with the DF through a decision tree of five splits has improved the diagnostic accuracy by 23\% regarding the single metrics.
\begin{figure}[t!]
	\centering
	\includegraphics[width=\linewidth]{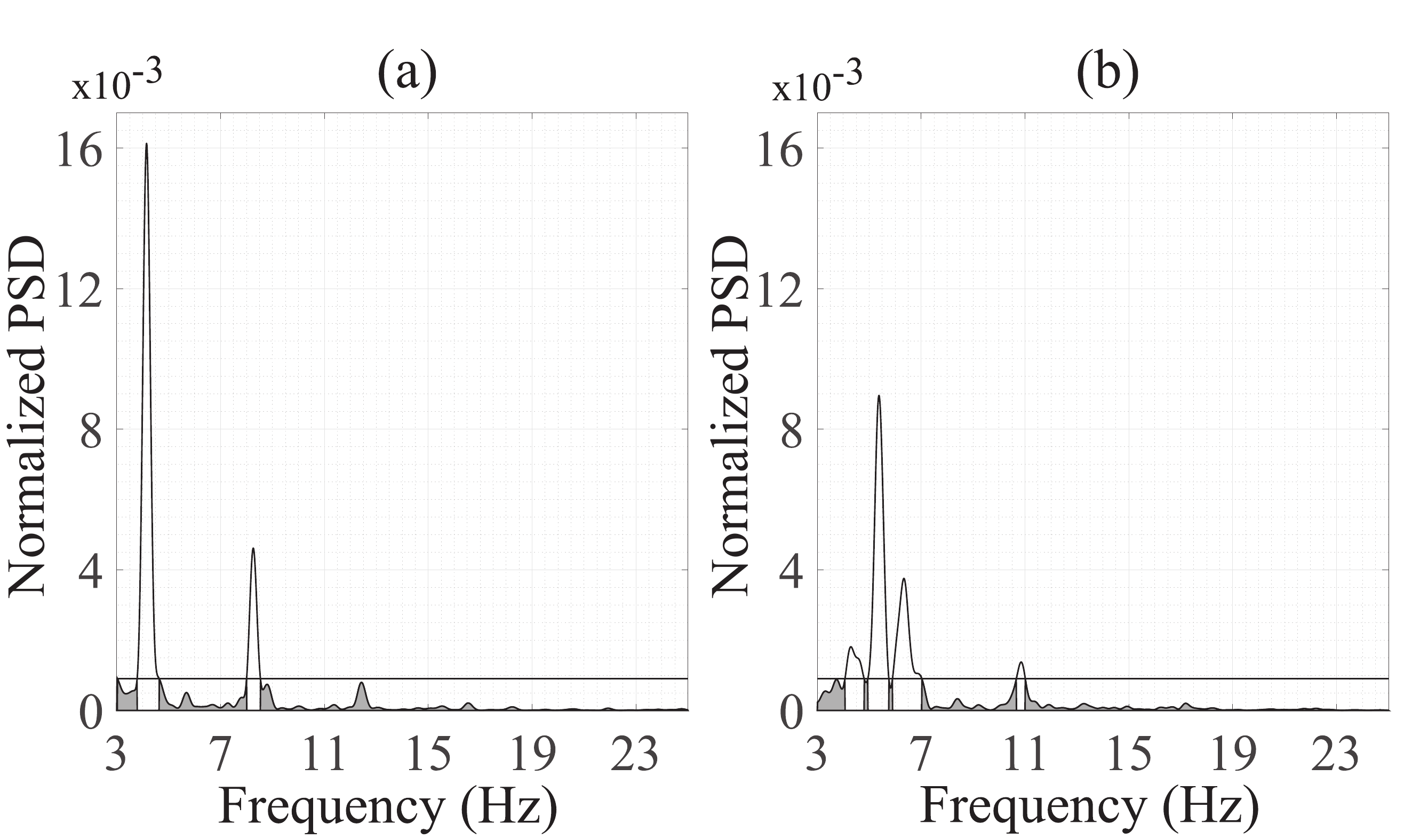}
	\caption{Example of PSD partition by $C_0$ complexity metric for a patient maintaining SR (a) and a patient who relapsed to AF (b).}
	\label{figure:1}
\end{figure}

The results obtained by the DF and nFWA agree with previous studies, which have observed that higher and lower values are respectively correlated with a higher probability of AF recurrence after CA~\cite{Nault:2009aa,Alcaraz2016}. In addition, the results of $C_0$ complexity indices has been associated with the degree of fibrillatory activity organization. Previous works have observed that patients with more organized fibrillatory activity, reﬂected by a greater harmonic weight compared to the rest of the frequency components, have a greater probability of responding successfully to any AF treatment, including CA \cite{Alcaraz2016}. Hence, as Figure~\ref{figure:1} (a) shows, the presence of strong DF and harmonic components in patients maintaining SR after CA follow-up period provided a higher irregular part of the PSD distribution, and therefore a higher $C_0$ complexity value. In contrast, patients that relapsed to AF (Figure~\ref{figure:1} (b)) presented a more disorganized PSD distribution with different low frequency components and not a clear main frequency, which entailed a lower value of $C_0$ complexity.
\par
All these results and observations suggest that spectral distribution of the $f$-waves can provide relevant information for the preoperative prediction of CA outcome. Nonetheless, this finding should be confirmed in the future on a wider set of patients. 
	

\section*{Acknowledgments}  
This research has received financial support from public grants PID2021-00X128525-IV0 and PID2021-123804OB-I00 of the Spanish Government, 10.13039/ 501100011033 jointly with the European Regional Development Fund (EU), SBPLY/17/180501/000411 and SBPLY/21/180501/000186 from Junta de Comunidades de Castilla-La Mancha, and AICO/2021/286 from Generalitat Valenciana. Moreover, Pilar Escribano holds a predoctoral scholarship 2020-PREDUCLM-15540, which is co-financed by the operating program of European Social Fund (ESF) 2014-2020 of Castilla-La Mancha.  

\bibliography{refs}

\begin{thebibliography}{10}
\expandafter\ifx\csname url\endcsname\relax
  \def\url#1{\texttt{#1}}\fi
\expandafter\ifx\csname urlprefix\endcsname\relax\def\urlprefix{URL }\fi

\bibitem{Zoni2014}
Zoni-Berisso M, Lercari F, Carazza T, Domenicucci S.
\newblock Epidemiology of atrial fibrillation: European perspective.
\newblock Clinical Epidemiology 2014;\hspace{0pt}213.

\bibitem{Hindricks:2021aa}
Hindricks G, et. al.
\newblock {2020 ESC Guidelines for the diagnosis and management of atrial
  fibrillation developed in collaboration with the European Association for
  Cardio-Thoracic Surgery (EACTS)}.
\newblock Eur Heart J 02 2021;\hspace{0pt}42(5):373--498.

\bibitem{Warmus2020}
Warmus P, Niedziela N, et~al.
\newblock Assessment of the manifestations of atrial fibrillation in patients
  with acute cerebral stroke - a single-center study based on 998 patients.
\newblock Neurol Res Jun 2020;\hspace{0pt}42(6):471--476.

\bibitem{Schotten:2016aa}
Schotten U, Dobrev D, et~al.
\newblock Current controversies in determining the main mechanisms of atrial
  fibrillation.
\newblock J Intern Med May 2016;\hspace{0pt}279(5):428--38.

\bibitem{Nattel:2014aa}
Nattel S, Guasch E, et~al.
\newblock Early management of atrial fibrillation to prevent cardiovascular
  complications.
\newblock Eur Heart J Jun 2014;\hspace{0pt}35(22):1448--56.

\bibitem{Schmidt2020}
Schmidt B, Brugada J, et~al.
\newblock Ablation strategies for different types of atrial fibrillation in
  {E}urope: results of the {ESC-EORP EHRA} atrial fibrillation ablation
  long-term registry.
\newblock Europace 2020;\hspace{0pt}22(4):558--566.

\bibitem{Walsh2018}
Walsh K, Marchlinski F.
\newblock Catheter ablation for atrial fibrillation: current patient selection
  and outcomes.
\newblock Expert Review of Cardiovascular Therapy
  2018;\hspace{0pt}16(9):679--692.

\bibitem{Dretzke:2020aa}
Dretzke J, Chuchu N, et~al.
\newblock Predicting recurrent atrial fibrillation after catheter ablation: a
  systematic review of prognostic models.
\newblock Europace 05 2020;\hspace{0pt}22(5):748--760.

\bibitem{Alcaraz2016}
Alcaraz R, Hornero F, Rieta JJ.
\newblock Electrocardiographic spectral features for long-term outcome
  prognosis of atrial fibrillation catheter ablation.
\newblock Annals of Biomedical Engineering 2016;\hspace{0pt}44(11):3307--3318.

\bibitem{Nault:2009aa}
Nault I, Lellouche N, et~al.
\newblock Clinical value of fibrillatory wave amplitude on surface {ECG} in
  patients with persistent atrial fibrillation.
\newblock J Interv Card Electrophysiol Oct 2009;\hspace{0pt}26(1):11--9.

\bibitem{Morin2016}
Morin DP, Bernard ML, et~al.
\newblock The state of the art: Atrial fibrillation epidemiology, prevention
  and treatment.
\newblock Mayo Clinic Proceedings 2016;\hspace{0pt}91(12):1778--1810.

\bibitem{Petr_nas_2018}
Petr{\.{e}}nas, et~al.
\newblock Lead systems and recording devices.
\newblock In Atrial Fibrillation from an Engineering Perspective. Springer
  International Publishing, 2018;\hspace{0pt} 25--48.

\bibitem{Leif2005}
S{\"o}rnmo L, Laguna P.
\newblock The electrocardiogram. a brief background.
\newblock In Bioelectrical Signal Processing in Cardiac and Neurological
  Applications. Elsevier, 2005;\hspace{0pt} 411--452.

\bibitem{Garcia:2018aa}
Garc{\'\i}a M, Mart{\'\i}nez M, R{\'o}denas, et~al.
\newblock A novel wavelet-based filtering strategy to remove powerline
  interference from electrocardiograms with atrial fibrillation.
\newblock Physiol Meas 11 2018;\hspace{0pt}39(11):115006.

\bibitem{Alcaraz:2008aa}
Alcaraz R, Rieta JJ.
\newblock Adaptive singular value cancelation of ventricular activity in
  single-lead atrial fibrillation electrocardiograms.
\newblock Physiol Meas Dec 2008;\hspace{0pt}29(12):1351--69.

\bibitem{Alcaraz:2011aa}
Alcaraz R, Hornero F, Rieta JJ.
\newblock Noninvasive time and frequency predictors of long-standing atrial
  fibrillation early recurrence after electrical cardioversion.
\newblock Pacing Clin Electrophysiol Oct 2011;\hspace{0pt}34(10):1241--50.

\bibitem{Yueli2008}
Lu Y, Jiang D, et. al.
\newblock Predict the neurological recovery under hypothermia after cardiac
  arrest using {C0} complexity measure of eeg signals.
\newblock In 30th Annual International Conference of the IEEE Engineering in
  Medicine and Biology Society. 2008;\hspace{0pt} 2133--2136.

\end{thebibliography}


  
  
      

\begin{correspondence}
Pilar Escribano Cano.\\
Postal address: Escuela Técnica Superior de Ingenieros Industriales de Albacete, Av. de España, s/n, 02071, Albacete, Spain.\\
E-mail address: pilar.escribano1@alu.uclm.es
\end{correspondence}

\end{document}